\documentstyle[twoside,fleqn,espcrc2]{article}

\newcommand{\AmS}{{\protect\the\textfont2
  A\kern-.1667em\lower.5ex\hbox{M}\kern-.125emS}}
\newcommand {\bea}{\begin{eqnarray}}
\newcommand {\eea}{\end{eqnarray}}
\newcommand {\be}{\begin{equation}}
\newcommand {\ee}{\end{equation}}

\title{
\begin{flushright}
\begin{small}
UPR-769-T,~hep-th/9708090\\
\end{small}
\end{flushright}
\vspace{1.cm}
Black Hole Horizons and the Thermodynamics of Strings\thanks{Based on talks 
presented by F.L. at SUSY'97, Philadelphia, May 27-31, 1997,
and by M.C. at STRINGS'97, Amsterdam, June 16-21, 1997.}}
\author{Mirjam Cveti\v{c} and Finn Larsen\\
David Rittenhouse Laboratories\\
University of Pennsylvania\\
Philadelphia, PA 19104}
       
\begin{document}

\begin{abstract}
We review the classical thermodynamics and the greybody factors 
of general (rotating) non-extreme black holes and
discuss universal features of their near-horizon geometry.
We motivate a microscopic interpretation of general black holes that
relates the thermodynamics of an effective string theory to the 
geometry of the black hole in the vicinity of both 
the outer and the inner event horizons. In this framework we
interpret several near-extreme examples, the universal low-energy
absorption cross-section, and the emission of higher partial waves 
from general black holes.
\end{abstract}

\maketitle

\section{Introduction}
In the last few years a precise correspondence between certain black holes and 
quantum states in string theory has been found. In particular, the 
degeneracies of bound states of D-branes have been calculated and 
for large charges the result agrees with the Bekenstein-Hawking area 
law for the black hole entropy~\cite{sv1}. 
The configurations that allow 
for the precise arguments are BPS saturated, i.e. they are 
supersymmetric black holes with zero temperature. It is interesting to 
consider also the near-BPS configurations because they exhibit non-trivial 
thermodynamical properties.  There is agreement between string 
theory and the macroscopic black hole properties in this
case~\cite{callan96a,strom96b} also,  but the reasoning is 
less securely founded in the microscopic theory. It is therefore important 
that the  counting of microscopic states can be supplemented with 
comparisons of dynamical processes~\cite{dmw,mathur,greybody}: the rate 
of Hawking decay and the precise energy dependence of the radiation 
in the semi-classical theory is in agreement with that in the string 
model. These results give strong support to the identification of the 
microscopic structure of black holes with an effective string 
also in the near-BPS case; the classical black holes ``look'' 
like strings.

The correspondence between the black holes and the  
effective strings is only possible when the semi-classical results take 
a very specific form. Thus, 
the logic can be turned around and we can use the semi-classical calculations 
as an efficient diagnostic test for the microscopic properties of the 
non-extreme black holes  which are currently inaccessible  by  other methods 
in string theory.
One might have expected that general black holes behaved very 
differently from 
strings but, as we shall see, even the general non-extreme black holes
``look'' surprisingly like strings. This does not imply that we already
understand the precise connection between an effective  string theory
and non-extreme black holes, but it suggests the existence 
of a precise description, and some of its features can already be 
discerned. 
The situation is reminiscent of the early works on the BPS 
case which took classical black holes as starting points and found 
that these solutions exhibited properties that were in striking agreement 
with expectations from the microscopic string 
theory~\cite{duff94,sen95,structure,cfthair1}. Subsequently, the D-brane 
description provided a foundation for those observations.

This contribution is based on  work 
presented in ~\cite{cy96b,fl97,cl97a,cl97b}. It is
organized as follows. In sec.~\ref{sec:title} it is explained how 
some features of an underlying string description can be read off 
directly from the classical geometry. In sec.~\ref{sec:counting} we 
make these ideas explicit in the context of the statistical mechanics of
the most general five dimensional black holes. 
We show that previously known results are recovered in the BPS and 
near-BPS limits; and we discuss some tests of the extension 
to the general non-extreme case. The four dimensional case is more
involved than the five-dimensional one but, as discussed 
in~\cite{cl97b}, most results nevertheless carry over with only
minor modifications.

As explained above 
the most detailed test of the correspondence between black holes and 
strings is that the spectrum of the Hawking radiation agrees in the two
descriptions. In sec.~\ref{sec:waveeq} we develop this approach by 
discussing the  features of the 
field equation of a minimally coupled scalar. In 
sec.~\ref{sec:greybody} we use these results to find 
the greybody factors of the black holes and highlight their string theory 
interpretation. This provides a test of the geometric picture
presented in sec.~\ref{sec:title}. In sec.~\ref{sec:conclusion} 
we summarize the successes and the limitations of the string description of 
non-extreme black holes. 

\section{Thermodynamics of Strings and Geometry of Black Holes}
\label{sec:title}
A characteristic feature of strings is that their spectrum divides into
two distinct sectors associated with the right (R) and left (L) moving 
excitations, respectively. The space of states therefore decomposes 
into a direct
product of two spaces. Consequently 
the entropy, calculated from the degeneracies 
of these states, is the sum of two contributions:
\begin{equation}
S=S_R+S_L.
\label{ents}
\end{equation} 
The condition  for this 
division to be meaningful is that the coupling between the R- and L-moving 
 modes
is much weaker than the coupling within each sector. Under the same condition
{\it two independent temperatures} can be introduced; $T_{R,L}$ for 
for the R- and L-moving  modes, respectively.
The temperature of the combined system is related to that of its
parts as: 
\begin{equation}
T^{-1}_H={1\over 2}(T_R^{-1}+T_L^{-1}). 
\label{temps}
\end{equation}
Other thermodynamic 
potentials similarly  split into R- and L-moving contributions:
there is one potential for each of the two kinds of modes, and a 
combined potential for the full system. In sum weakly coupled string 
theory gives rise to two distinct sets of thermodynamic variables. 

The thermodynamics of black holes has a  different starting point than 
the statistical mechanics of strings described above; the thermodynamic 
variables are identified with geometrical features of space-time rather 
than specific microscopic states. For example the  Bekenstein-Hawking (BH)
entropy is related to the area of the outer event horizon $A_{+}$ as:
\be
S={A_{+}\over 4G_N}~,
\label{ent}
\ee
and the  Hawking temperature is related to the surface acceleration at the 
outer event horizon $\kappa_+$ as:
\be
T_{H}={\kappa_{+}\over 2\pi}~.
\label{temp}\ee
Here $G_N$ is the Newton's constant.

In a microscopic description of black holes as weakly coupled effective 
strings the geometrically defined thermodynamic variables (\ref{ent}) 
and (\ref{temp}) must each be divided into two terms, as in the string 
system (\ref{ents})  and (\ref{temps}). The general rotating black holes 
of toroidally compactified string theory strongly suggest how this
should be done~\cite{cy96b}, as we explain in the subsequent section. 
Here we present our proposal in geometrical terms, rendering it 
independent of the specific dimensionality of space-time and the 
compactification of the string theory. It is also invariant under 
coordinate changes as well as duality transformations. 

The idea is that
general black holes have {\it two sets of thermodynamic variables}  
associated with the {\it  two event horizons}, the inner
and the outer one. Specifically, the R-- and L--moving entropies are:
\be
S_{R,L}= {1\over 2} ( {A_{+}\over 4G_N}\mp {A_{-}\over 4G_N})~,
\label{eq:rlent}
\ee
where $A_\pm$ are the areas of the inner and outer event horizons;
and the R-- and L--moving temperatures:
\be
{1\over T_{R,L}} = {2\pi\over\kappa_{+}}\pm {2\pi\over\kappa_{-}}~,
\label{eq:rltemp}
\ee
where $\kappa_\pm$ are the surface accelerations of the inner and outer event
horizons. These geometric definitions of the temperatures are consistent
with the result of the first law of thermodynamics applied to the R-- and
L--moving  entropies independently~\cite{cl97a}:
$T_{R,L}^{-1} =2({\partial S_{R,L}\over \partial M})_{\vec{Q},\vec{J}}~
$.

It is an unfamiliar idea that the inner event horizon plays 
any role at all, as it is effectively isolated from an outside 
observer. The ultimate justification of this idea comes from its 
applications; the calculations presented in the following sections 
indeed support that, at least formally, the ``contributions'' from the 
inner and the outer horizon appear on equal footing.

\section{General Black Holes in String Theory --- 
the Five-Dimensional Example}
\label{sec:counting}
As the  working example we consider the most general five-dimensional
black hole in toroidally compactified string theory. It depends on
the ADM mass, $M$, three $U(1)$ charges $Q_i$, and two angular momenta 
$J_{R,L}$. We parameterize these physical variables by
the non-extremality parameter $\mu$, the three boost parameters $\delta_i$,
and the (bare) angular momentum parameters $l_{1,2}$ introduced through:
\bea
M &=& {\textstyle{1\over 2}}\mu\sum_{i=1}^3 \cosh 2\delta_i~,\\
Q_i &=&{\textstyle {1\over 2}}\mu \sinh 2\delta_i~~~;~i=1,2,3~,\\
J_{R,L}&=& {\textstyle{1\over 2}}\mu (l_1\pm l_2)
(\prod_{i=1}^3 \cosh\delta_i \mp
\prod_{i=1}^3 \sinh\delta_i) \label{eq:angmom}
\eea
Units are such that the gravitational coupling constant 
in five dimensions is:
$
G_5 = {\textstyle{\pi\over 4}}~{(\alpha^\prime)^4 g^2/( R_1 R_2 R_3 R_4 R_5)}
={\textstyle{\pi\over 4}}~,
$
where $R_i$ are the radii of the compact tori, the string coupling is 
normalized so that  under S-duality  $g\rightarrow g^{-1}$, and
$\alpha^\prime$ is the Regge slope. 
The three $U(1)$ charges of the generating solution can be chosen to be 
coupled to D1-branes, D5-branes, and Kaluza-Klein charge, respectively. 
In this
case they are related to quantized (integral) charges $n_i$ through:
\bea
Q_1 &=& {n_1 R_1\over g\alpha^\prime}~~~~~~~~~~~~~~~~~~{\rm (D1-branes)}  \\
Q_2 &=& {n_2 R_1 R_2 R_3 R_4 R_5\over g (\alpha^\prime)^3} 
~~~{\rm (D5-branes)} \\
Q_3 &=& {n_3\over R_1}~~~~~~~~~~~~~~~~~~~~{\rm (KK-charge)}
\eea
These assignments will be assumed for definiteness, but  
many other choices are equivalent by duality~\footnote{In fact the 
explicit generating solution given in~\cite{cy96a} employs  
three charges in the NS-NS sector: electric winding and momentum charges, 
and a charge associated with the  dualized anti-symmetric tensor 
field.}. Indeed, duality 
transformations in the maximal compact subgroup of the full duality 
group generate the most general black hole~\cite{hull96}  when they act on the 
generating solution defined by (7--9)~\footnote{For a review of general black
hole solutions in toroidally compactified string theory see: \cite{c96}.}.
(This procedure was made explicit 
in~\cite{gaida96}.)

 From the explicitly known solutions the areas of the inner and outer
horizons can be read off~\cite{cy96a,cy96b}. The entropies calculated 
using (\ref{eq:rlent}) become:
\be
S_{R,L} = 2\pi\sqrt{\textstyle{{1\over 4}}\mu^3 
(\prod_i \cosh\delta_i\mp\prod_i \sinh\delta_i)^2-J^2_{R,L}}
\label{eq:srl}
\ee
According to our interpretation these expressions should be identified 
with the entropies of the R-- and L--moving modes of the underlying string 
theory. 

The R-- and L--moving temperatures are similarly calculated from the surface 
accelerations at the inner and outer event 
horizons, using eq. (\ref{eq:rltemp}). They are:
\be
{1\over T_{R,L}}= 
{\pi\mu^2 (\prod_i \cosh^2\delta_i-\prod_i \sinh^2\delta_i)\over
\sqrt{{1\over 4}\mu^3 
(\prod_i \cosh\delta_i\mp\prod_i \sinh\delta_i)^2-J^2_{R,L}}}
\label{eq:trl}
\ee
The physical content of these formulae becomes clearer in various 
limiting cases that we consider in the following.

\paragraph{The BPS limit:}
The extremal case corresponds to the limit where 
$\delta_i\rightarrow\infty$. This limit is only regular when also 
$J_R\rightarrow 0$. The BPS mass is given by the sum of the three charges:
\be
M= Q_1 + Q_2 + Q_3~;
\ee
so the black hole can be interpreted as a marginal bound state of three 
kinds of objects. The degeneracy of this composite object is given by 
the (exponential of) the entropy~\cite{rotation1}:
\be
S= 2\pi\sqrt{Q_1 Q_2 Q_3 - J_L^2}= 2\pi\sqrt{n_1 n_2 n_3 - J_L^2}~.
\label{eq:microent}
\ee
In the intermediate step the quantization conditions~(10--12) on the charges
were used. The moduli cancel out~\cite{structure,cfthair1,kallosh96b} so 
that the entropy can be interpreted directly in terms of the underlying 
constituents. Note that $S_R=0$ and $S=S_L$; so in the BPS-limit
the black hole entropy does not divide into two terms. 
According to (\ref{eq:rlent}) the R--moving contribution 
indeed vanishes in 
the geometric interpretation because the two horizons coincide.

The string theory calculation that leads to (\ref{eq:microent}) takes
as its starting point the superconformal field theory (SCFT)  with the target
space~\cite{sv1}:
\be
C =(T^4)^{n_1 n_2}/\Sigma_{n_1 n_2} ~,
\ee
and level $n_3$.  Here $\Sigma_{k}$ is the permutation group of $k$ objects.
It acts on the product manifold in
an orbifold construction and introduces twisted sectors that contribute
fractionally to the momentum~\cite{mathur96,fatbh,dmvv}. 
In the black hole limit it is the sector 
with the maximal fractionation that provides the most important 
contribution. The contribution of this sector can be captured by the 
effective level:
\be
N_L =n_1 n_2 n_3 - J_L^2~, 
\label{eq:nl}
\ee
of a superstring with target space $T^4$, or more generally a SCFT
with  the central charge $c=6$ (${\hat c}=4$). Indeed, 
(\ref{eq:microent}) is recovered 
using $S\simeq 2\pi\sqrt{{c\over 6}N}$, valid at large $N$.
Although the concept of an effective level is approximate in general it 
becomes precise in the black hole regime. It is useful because it captures 
the symmetry between the three charges required by duality.

The angular momentum is the $U(1)$ component of 
the local $SU(2)$ world-sheet current of the $N=4$ 
SCFT~\cite{rotation1}. The projection onto the sector with a specific 
angular momentum is multiplication 
by an operator with the appropriate $U(1)$ world-sheet charge, and the 
scaling 
dimension (conformal weight) of this operator is responsible for the 
effective subtraction of $J_L^2$ in eq.  (\ref{eq:nl}).

\paragraph{Extreme Kerr-Newman limit:}
A related analysis can be applied to the extreme Kerr-Newman type 
black hole solutions. This limit is 
achieved by taking $(l_1-l_2)^2\rightarrow \mu$. Then $S_R=0$, again,  
and thus $S=S_L$ which takes the form:
\begin{equation}
S=2\pi\sqrt{n_1 n_2 n_3 +J_R^2-J_L^2}~.
\label{eq:microentp}
\end{equation}
Although this solution is far from the BPS-limit the BH entropy is 
again moduli independent; and the  microscopic  entropy can be 
modeled~\cite{hlm} by an effective string  with $c_R=c_L=6$. The 
R-moving sector contributes to the angular momentum component 
$J_R$ but the excitation level is $N_R=0$.  

\paragraph{The dilute gas limit:}
In this limit two of the boosts are large, say,  $\delta_{1,2} \gg 1$
~\cite{strom96b,greybody}. The corresponding two charges are 
conventionally chosen as the D1-brane and D5-brane charges. 
These charges act as backgrounds while the third charge signifies a
deviation from the BPS-limit: it couples to both R-- and L--moving momentum 
carrying waves. The effective levels can be 
found from the excess energy over the BPS-limit, assuming that the 
background branes fractionate the contributions to the levels, as 
in the BPS-limit, and that the important degrees of freedom are 
excited versions of the effective $c=6$ string that appear in the 
BPS-limit:
\be
N_{R,L}= Q_1 Q_2 ~{\textstyle {1\over 4}}\mu e^{\mp 2\delta_3}-J^2_{R,L}~.
\ee
The corresponding entropies indeed agree with eq.  (\ref{eq:srl}) in the 
limit where two boosts are large. 
This verifies the role of the inner horizon in the dilute gas limit.

\paragraph{One large boost:}
The case where only one of the boosts is large, say,  $\delta_1\gg 1$,
can be modeled similarly~\cite{maldacena96b,mathur97}: the charge that
corresponds to the large boost acts as a background that is
inert, except that it induces the fractionation. The excess energy,
$\Delta M$,
is distributed  among the  states in the  spectrum in a  way that is similar
to  the  fundamental string with both momentum and winding charge.
Considering
first the standard relation of perturbative string theory:
\be
N_{R,L}^{\rm pert} = \Delta M^2 - (Q_2\pm Q_3)^2 = 
\mu^2 \cosh^2 (\delta_2 \mp \delta_3)
\ee
We also take into account the fractionation and the projection onto a
specific angular momentum sector. Then the effective levels become:
\be
N_{R,L}= Q_1 ~\mu^2 \cosh^2 (\delta_2 \mp \delta_3)-J^2_{R,L}~.
\ee
The corresponding entropies reproduce eq.  (\ref{eq:srl}) in the limit
where one boost is large, thus   verifying  the role of the inner horizon 
in the regime of one large boost.

Note that this case includes general non-extreme black 
holes in the infinite momentum frame. This may be  relevant for the 
description of black holes in the framework of M(atrix)-theory~\cite{bfss}.

\paragraph{The general case:}
The general expression for the  BH entropy (\ref{eq:srl}) can be 
accounted for quantitatively by a non-critical string with central 
charge $c_R=c_L=6$, excited to the 
effective levels:
\be
N_{R,L}= \textstyle{{1\over 4}}\mu^3 
(\prod_i \cosh\delta_i\mp\prod_i \sinh\delta_i)^2-J^2_{R,L}
\label{eq:nrl}
\ee
An important test of this idea is that it reduces to the limits 
described above when an appropriate number of boosts $\delta_i$ are 
large.  However, in the general case the mass spectrum is not known 
from other considerations. Thus, unlike the previous special cases we 
cannot use this knowledge to justify the effective levels 
(\ref{eq:nrl}). Instead 
our interpretation provides the spectrum in the black hole regime. 

As an alternative test of the proposal use $Q_1 Q_2 Q_3 = n_1 n_2 n_3$
to find:
\be
N_R - N_L = n_1 n_2 n_3 +J_L^2 - J_R^2~.
\ee
Since the angular momenta have been normalized so that $J^2_R-J^2_L$ is
quantized as integers, the expression above
is independent of moduli and {\it  it is  an integer}. 
This result  lends support to the hypothesis that $N_{R,L}$ are 
individually quantized.
 
The length of the effective string can be derived from thermodynamics
alone. Indeed, non-interacting gases in one dimension satisfy:
\be
S = {\pi c\over 6}~T{\cal L}~,
\ee
where ${\cal L}$ is the ``volume'' --- length --- of the gas.  
Thus, using (\ref{eq:srl}) for $S$ and (\ref{eq:trl}) for $T$,
the length  $\cal L$ of the effective string 
becomes\footnote{In~\cite{cl97b} this result was inferred from the
absorption cross-section. The derivation of ${\cal L}$ given here was 
found independently in~\cite{kastor}.}:
\be
{\cal L} = 2\pi\mu^2
(\prod_i \cosh^2\delta_i-\prod_i \sinh^2\delta_i)~. 
\label{stlen}
\ee
This length scale is {\it independent of the
angular momenta}, which provides another test of the model. Namely, in
string theory angular momenta are 
implemented as projections on the Hilbert space of states and 
do not affect the length of the string. Moreover, it is satisfying 
that the same string length $\cal L$  is found in the R-- and L--moving sectors.
 ${\cal L}$ reduces to ${\cal L}=2\pi n_1 n_2 R$
in the BPS-limit but the general expression  (\ref{stlen}) 
is also valid for non-BPS
black holes. The fact that the effective string length $\cal L$ 
increases with 
the size of the black hole is potentially important for the issue
of information loss. Indeed, for large black holes the scale
${\cal L}$ is much larger than the Planck scale, where
string effects are usually assumed to become important.

These considerations also apply to the static cases with one or more  
$\delta_i=0$, including the Schwarzschild solution with all $\delta_i=0$.
However, in these examples $S_R=S_L$ and $T_R=T_L=T_H$; so the 
characteristic string features of the thermodynamics are absent. 
In the geometric 
interpretation this is a consequence of the degenerate limit of the vanishing 
inner horizon area. However, this presumably does not imply any limitation
in the string description. A plausible interpretation is simply that,
in this case, there is an equilibrium between the two sectors.

\section{The Wave Equation --- Minimally Coupled Scalar Field}
\label{sec:waveeq}
A detailed test of the correspondence between black holes and strings 
is the 
spectrum of Hawking radiation. As preparation for this calculation
we consider general properties of the wave equation for a minimally 
coupled scalar field:
\be
{1\over\sqrt{-g}}\partial_\mu (\sqrt{-g}g^{\mu\nu}\partial_\nu \Phi) =0~.
\ee
It is straightforward (but tedious) to insert the metric given 
in~\cite{cy96a}. The equation turns out to be separable.  We can therefore
write the wave function as:
\be
\Phi\equiv 
\Phi_0(r)~\chi(\theta)~e^{-i\omega t+im_R(\phi+\psi)+im_L(\phi-\psi)}~,
\ee
where $\phi$ and $\psi$ are the two  azimuthal angles. We also introduce a 
dimensionless variable $x$ that is related to the standard 
radial coordinate $r$ through:
\be
x \equiv {r^2 - {1\over 2}(r^2_{+}+r^2_{-})\over
(r^2_{+}-r^2_{-})}~.
\label{eq:xdef}
\ee
In this coordinate system the outer and inner event horizons  
are at $x={1\over 2}$ and $x=-{1\over 2}$, respectively, and the 
asymptotically flat region is at $x=\infty$. Then the radial part of
the wave equation becomes~\cite{cl97a}:
\bea
&~&\partial_x(x^2-\textstyle{{1\over 4}})\partial_x\Phi_0
+\textstyle{{1\over 4}}[x\Delta\omega^2+M\omega^2-\Lambda+
\nonumber
\\
&+&{1\over x-{1\over 2}}
({\omega\over\kappa_{+}}-m_R {\Omega^R\over\kappa_{+}}
-m_L {\Omega^L\over\kappa_{+}})^2
\label{eq:geneq}
\\
&-&{1\over x+{1\over 2}}({\omega\over\kappa_{-}}-m_R {\Omega^R\over\kappa_{+}}
+m_L {\Omega^L\over\kappa_{+}})^2]\Phi_0 = 0~. \nonumber
\eea
In spite of its generality this equation is no more complicated than 
similar ones that have 
been considered in various special 
cases~\cite{greybody,mathur97,strominger97a}.

We  first discuss the various terms  in eq. (\ref{eq:geneq}). 
The variable $\Delta$ is defined by 
$\Delta=r^2_{+}-r^2_{-}$; so at very large $x$ the equation 
becomes:
\be
({1\over r^3}{\partial\over\partial r}r^3 {\partial\over\partial r}
+\omega^2)\Phi_0 = 0~.
\ee
This is the radial part of the Klein-Gordon equation in flat
space. Thus, the term ${1\over 4}x\Delta\omega^2$ is simply the 
energy of the perturbation in the absence of the black hole.

The term ${1\over 4}M\omega^2$ can be interpreted as 
the Coulomb-type screening due to the gravitational field. 
At large $x$ this term is suppressed relative to flat space by one power 
of $x\sim r^2$ as expected for a Coulomb potential in five dimensions.

The variable $\Lambda$ is the eigenvalue of the angular Laplacian. It 
takes the form:
\be
\Lambda = n(n+2) +
{\cal O}(\omega^2)~,
\ee
where the corrections  ${\cal O}(\omega^2)$ 
are due to the rotation of the background and  
are discussed in~\cite{cl97a}. This term is also suppressed by one power 
of $x\sim r^2$, as expected for an angular momentum barrier. 

The terms considered so far are manifestations of the long range fields and 
the flat asymptotic space. The 
remaining two terms diverge at the outer   ($x={1\over 2}$) and inner 
($x=-{1\over 2}$) horizons, respectively, and so are specific for black 
hole backgrounds.  The modes close to 
the outer horizon become (taking $m_R=m_L=0$, and thus ignoring the effects of
angular velocities, $\Omega^{R,L}$, for simplicity):
\be
\Phi_0 \sim (x-{\textstyle{1\over 2}})^{\mp{i\omega\over 2\kappa_{+}}}
e^{-i\omega t} ~.
\ee
The branch-cut around $x={1\over 2}$ is tantamount to the existence
of Hawking radiation with temperature $T_H = {\kappa_{+}\over 2\pi}$, a
well known  result found in several different ways in the seventies.

The modes close to the inner horizon similarly become:
\be
\Phi_0 \sim (x+{\textstyle{1\over 2}})^{\mp{i\omega\over 2\kappa_{-}}}
e^{-i\omega t}~.
\ee
The significance of these modes is less clear because the inner horizon 
is not expected to have any effects on the asymptotic observers. However, 
there is a striking parallel between the two horizons.
It is therefore natural to suspect that the branch-cut around $x=-{1\over 2}$
nevertheless has a thermal interpretation. As explained in 
sec.~\ref{sec:title} the inner horizon temperature
$T_{-}={\kappa_{-}\over 2\pi}$ combines with the Hawking 
temperature, $T_H$, and forms two new temperatures 
$T^{-1}_{R,L}=T^{-1}_H\pm T^{-1}_{-}$ which  can be attributed
the R-- and L--moving string modes, respectively. 

\paragraph{Hidden Supersymmetry:}
For rotating black holes there are not sufficiently many conserved 
quantities that the separation of variables can be guaranteed.
It is therefore a surprise that we were able to do so\footnote{We would
like to thank G. Gibbons for pointing this out.}. 
An analogous surprise is well known in the 
context of four-dimensional Kerr-Newman black holes. There the separation of
variables is a consequence of conserved fermionic charges that are 
related to the 
more familiar conserved bosonic charges by the supersymmetry 
algebra~\cite{susysky}. It is believed that the significance of the
supersymmetry in this context is  different from its role in particle 
physics. However, it would be interesting to reexamine this question 
in light of the relation between black holes and superstrings.

\paragraph{String symmetries:}
The wave equation in the region close to the horizons has a simple
form that may be universal, as the identical equation appears in both 
four and five dimensions~\cite{cl97b}. We now express this structure 
mathematically. 

We introduce the 
dimensionless Rindler time $\tau$ that is a regular time-like coordinate 
close to the outer horizon. The monodromy around the coordinate
singularity is encoded by an imaginary period ${2\pi iT_H\over\omega}$. 
Analogously, we introduce an ``inner-horizon Rindler time'' $\sigma$. 
This coordinate is space-like, as the signature is changed close to 
the inner horizon, and it has an imaginary period 
${2\pi iT_{-}\over\omega}$. With these auxiliary 
variables the radial equation (\ref{eq:geneq})
(without  the flat space term $ {1\over 4}x\Delta\omega^2$)
becomes the eigenvalue equation of the operator:
\be
{\cal H}_r = 
-\partial_x (x^2 - \textstyle{{1\over 4}})\partial_x
-{1\over x-{1\over 2}}\partial^2_\tau
+{1\over x+{1\over 2}}\partial^2_\sigma
\label{eq:hrdef}
\ee
with the eigenvalue ${1\over 4}(\Lambda-M\omega^2)\equiv h(h-1)$. 
The significance of the variable $h$ (as a conformal dimension of an 
effective string vertex)  will be explained in the next section.
The operator (\ref{eq:hrdef}) is the quadratic Casimir of the group 
$SL(2,{\bf R})_R\times SL(2,{\bf R})_L$. Its appearance shows that the  
propagation in 
the black hole background is in fact the motion on this group manifold. 
A similar description has previously appeared in the context of exact 
conformal field theories that describe two-dimensional 
black holes, but the  relation with the present results
is not clear (see {\it e.g.,}~\cite{dvv92}).

The compact generators $R_3$ and $L_3$ of the two 
$SL(2,{\bf R})_{R,L}$ 
groups are diagonal and their eigenvalues are:
\bea
R_3 &=&\textstyle{{1\over 2}} (\partial_\tau + \partial_\sigma)= 
{\omega\over 4\pi T_R}~, \\
L_3 &=& \textstyle{{1\over 2}}(\partial_\tau - \partial_\sigma)= 
{\omega\over 4\pi T_L}~.
\eea
Note that the natural variables of the group are the sum and the difference 
of $\tau$ and $\sigma$, rather than $\tau$ and $\sigma$ themselves. 
The auxiliary variables $\tau$ and $\sigma$ are
localized ``times'' close to each of the two horizons. Thus,
in a definite sense, it is the sum and the difference of the two horizons that 
are singled out by the group structure.  This result is satisfying because 
it is precisely these combinations of the surface accelerations 
that we assign microscopic significance.

The wave equation exhibits an obvious symmetry between inner and outer 
horizon terms that can be expressed in terms of the group generators 
as an automorphism of the algebra that takes $R_3\to R_3$ and 
$L_3\to -L_3$. In string theory the T-duality symmetry acts in precisely 
this way.

It is intriguing that symmetries that are closely associated with string 
theory are realized explicitly through the wave equation in the general 
black hole  background. 
However, we must emphasize their precise significance remains 
unclear.

\section{The Greybody Factors}
\label{sec:greybody}
We  now consider the solutions of the wave equation discussed in 
sec.~\ref{sec:waveeq}, 
following~\cite{greybody,mathur97,strominger97a,cl97a}

At large $x$ we consider the radial equation in flat space and we include
the effects of the long range fields from the black hole. The solution
to this approximate equation is a Bessel function. Next we consider 
the equation with only the unperturbed energy 
${1\over 4}\Delta x\omega^2$ omitted. Then the solution is a hypergeometric 
function. 
These wave functions with validity in some region of space can,
under conditions discussed later, be combined to form an approximate 
solution that is valid throughout.  The character of the result 
depends on the value of the potential terms 
${1\over 4}(M\omega^2-\Lambda)=h(h-1)$ in the region where the two 
partial solutions are both valid. 

\paragraph{Matching on zero potential:}
We first consider the situation where the potential vanishes in 
the ``matching'' region, i.e. $h=1$. In this case the S-wave 
absorption cross-section becomes~\cite{mathur,greybody,cl97a}:
\be 
\sigma_{\rm abs}^{(0)}(\omega)= {\pi^2 {\cal L}\omega\over 2}
{(e^{\omega/T_H}-1)
\over (e^{\omega\over 2T_L}-1)(e^{\omega\over 2T_R}-1)}~.
\label{eq:sabs}
\ee
Using the formula $S T_H = 2\pi{\cal L}T_R T_L$ it can be shown that
the absorption cross-section approaches the area of the outer horizon ($A_+$)
 as 
$\omega\rightarrow 0$. This is the universal low-energy
result~\cite{gibbons96} (which  also applies to rotating
black holes~\cite{cl97a}.) 
 
 From the absorption cross-section the 
emission rate follows
using detailed balance:
\bea
\Gamma_{\rm em}^{(0)}(\omega) &=&\sigma_{\rm abs}(\omega)~
{1\over e^{\omega/T_H}-1}~{d^4 k\over (2\pi)^4} 
\label{eq:emrate}
\\
&=& {\pi^2 {\cal L}\omega\over 2}{1\over (e^{\omega\over 2T_L}-1)
(e^{\omega\over 2T_R}-1)}~{d^4 k\over (2\pi)^4} \nonumber 
\eea
It is important to note that the Bose distribution with the
Hawking temperature canceled out. Therefore the final result depends 
only on the quantities $T_{R,L}$ and ${\cal L}$ that have significance in
the effective string description. 
In fact there is an explicit microscopic interpretation
of this formula: the emission is the result of a 
two-body process~\cite{callan96a,mathur,greybody,cl97a}. The Bose 
distributions are phase space factors of the R-- and L--moving 
quanta propagating on the string. Moreover, the amplitude for the
annihilation of
two quanta colliding head-to-head on a string of length ${\cal L}$ 
can be calculated, using only the Nambu-Goto form of the string action. 
The result of this calculation is identical 
to eq. (\ref{eq:emrate}). Thus the 
agreement between the space-time calculation and the microscopic 
interpretation involves the functional dependence on the 
energy, and all numerical factors agree as well.

The microscopic two-body interpretation of the emission can be employed
as a model whenever the form of the semi-classical result is of the type
(\ref{eq:sabs}). A sufficient condition for this is the low 
energy requirement $M\omega^2 \ll 1$
~\footnote{In~\cite{cl97a} two additional constraints were given.
They are however automatically satisfied, given this condition.}. 
This implies $(r^2_{+}-r^2_{-})\omega^2\ll 1$; 
thus the wave length of the probe is so large that the target cannot be 
positively identified as a black hole. On the other hand, the condition also 
implies ${\cal L}\omega^4\ll 1$. This presumably  implies that the target
cannot be unambiguously identified as a string either.

(i) An important special case is the dilute gas limit~\cite{greybody}, 
defined in sec.~\ref{sec:counting} as  
$\delta_{1,2}\gg 1$. In this case the low-energy condition
$M\omega^2 \ll 1$ is satisfied for 
frequencies $\omega\sim T_R \sim T_L$. Therefore the calculation is sensitive 
to the Bose distribution factors in eq.  (\ref{eq:emrate}). This verifies
in detail that the string temperatures have been correctly identified in 
the dilute gas regime.

(ii) Another interesting example is the  regime of 
rapidly spinning black holes~\cite{cl97a}, which 
is obtained by tuning the bare angular momenta $l_{1,2}$, defined
in eq. (\ref{eq:angmom}), so that
$l_2=0$ and $\mu-l^2_1=\mu\epsilon^2\ll\mu$.
As in the dilute gas case, the low energy condition is satisfied for 
$\omega\sim T_R \sim T_L$ so the Bose factors are significant. However,
now  there are {\it no conditions on the boosts} $\delta_i$; so
the functional dependence of the temperatures $T_{L,R}$ 
 and the string length $\cal L$
 on all three   boosts 
$\delta_i$ is tested in detail.

(iii) As a final example, consider the limit of very low energies where
the universal absorption cross-section $\sigma_{\rm abs}^{(0)}=A_{+}$ 
is valid for all black holes. The two-body model still
applies~\cite{susskind96} but the test afforded by the calculation 
of the emission rates is weaker because it does not involve the  
functional dependence on $\omega$. However, the agreement still involves 
the dependence on the independent black hole parameters 
$\mu$, $\delta_i$, and $l_{1,2}$. This result thus provides evidence 
that, at low energies, {\it the effective string model applies to 
all black holes}.

\paragraph{Matching on non-zero potential:}
The two-body annihilations considered so far are the simplest 
decay processes, but more complicated ones give important additional
information. Recall that one of the conditions for the validity of 
the two-body form of the emission rate is that the potential vanishes in
the region where the asymptotic solution and the near horizon solution
are matched. This condition can be replaced 
with the milder assumption that the two solutions can be matched in a 
region where the potential is any 
constant~\cite{mathur97,strominger97a,cl97a}.
In this case the final result for the emission rate into the n-th partial 
wave becomes:
\bea
\Gamma_{\rm em}^{(n)}(\omega) 
&=&{4\pi(n+1)^2 \over\omega^3 \Gamma^2(2h)} 
\left({{\cal L}\omega^2\over 8\pi}\right)^{2h-1}\times 
\label{eq:emrategen} \\
&\times & G^h_{T_R}\left({\omega\over 2}\right)~G^h_{T_L}\left({\omega\over 2}
\right)
~{d^4 k\over (2\pi)^4}~.
\nonumber 
\eea
where:
\be
G^h_{T}\left({\omega\over 2}\right)\equiv (2\pi T)^{2h-1}
e^{-{\omega\over 4T}} 
{|\Gamma (h- {i\omega\over 4\pi T})|^2\over \Gamma(2h-1)}.
\label{eq:green}
\ee
In the effective string  description $G^h_{T_{R,L}}({\omega\over 2})$ has an
interpretation as  the Fourier transform of the 
canonically normalized thermal Green's functions 
for a conformal field with the scaling dimensions  
$h_R=h_L=h$~\cite{gubser97b}:
\be
G^h_{T_R} (z) = \left({\pi T_R
\over\sinh \pi T_R z}\right)^{2h}
~,
\ee
and similarly  for $G^h_{T_L}$ with 
$R\rightarrow L$ and $z\rightarrow {\bar z}$.
When $h=1$ the general result (\ref{eq:emrategen}) reduces to the two-body
case  (\ref{eq:emrate}). For arbitrary $h$ the emission rate can 
still be interpreted in the framework of an effective $c_R=c_L=6$  SCFT
with $(4,4)$ supersymmetry that also accounts for the 
entropy~\cite{strominger97a}. In this model the operator that is 
responsible for the emission has dimension $h_R=h_L=h$ so that it 
accounts for the Green's functions in the emission rate. In this way 
the eigenvalue $h(h-1)$ of 
the quadratic Casimir of the $SL(2,{\bf R} )_R\times SL(2,{\bf R})_L$ 
group is related 
to the conformal dimension in the  effective string interpretation.

The general conditions for validity of the emission rate 
(\ref{eq:emrategen}) are given in ~\cite{cl97a}. 
Here we consider the sufficient condition that the partial wave number 
$n$ satisfies $M\omega^2\ll n^2$. For the various black holes with 
$M\omega^2\ll 1$, considered above, this holds for all $n$. 
More importantly, for typical frequencies $\omega\sim T_R\sim T_L$ 
the condition is valid for {\it arbitrary} black holes, as long 
as $n\gg 1$. The condition $M\omega^2\ll n^2$ can be written in terms 
of the impact parameter $b$ as $M\ll b^2$. Thus the probe sees the entire 
black hole as 
one entity. However, this does not imply that the black hole appears 
point-like; indeed, it ``looks'' like a string.

The condition $M\omega^2\ll n^2$ gives $h={n\over 2}+1$ so that the 
emission vertex operator has a free conformal field theory realization of the
schematic 
form:
\be
V\sim ~:\partial X(z)\bar{\partial} X(\bar{z}) 
[S^a (z) ]^{n}~[\bar{S}^{\dot a}(\bar{z})]^{n}:~,
\label{eq:vertex}
\ee
where  $:\ :$ denotes the 
normal ordering of the operators,  $X$ are coordinates of the string
 in the internal directions, and 
$S^a, {\bar S}^{\dot a}$ are  world-sheet fields  with conformal dimensions 
$(h_R,h_L)=(1/2,0)$ and  $(h_R,h_L)=(0,1/2)$, respectively,  
and whose indices  $(a, {\dot a})$ specify  
 quantum numbers of
the space-time spinors~\footnote{
The world-sheet  fields $s^a,{\bar s}^{\dot a}$ that transform
as spinors under the $SO(4)\sim SU(2)_R\times SU(2)_L$  
current algebra have the respective conformal dimensions  
$(1/4,0)$ and  $(0,1/4)$. Thus we can write
$S^a\sim s^a\chi, {\bar S}^{\dot a}\sim {\bar s}^{\dot a} {\bar \chi}$
where the fields $\chi,\bar \chi$ have conformal dimensions 
$(1/4,0)$ and $(0,1/4)$. Presumably $\chi, \ {\bar \chi}$ are 
associated with the internal 
degrees of freedom and might reflect the 
 $SL(2,{\bf R})_R\times SL(2,{\bf R})_L$
symmetry of the scattering equation.}.
 The emission is therefore
interpreted as a many-body process that involves $1$ boson and $n$ 
fermions in both R-- and L--moving sectors. Lorentz invariance
implies that the coupling of this vertex operator to the outgoing
field involves $n$ derivatives acting on the
outgoing field. This gives rise to a factor of $\omega^{2n}$ in the
rate and, remembering the normalization $\omega^{-1}$ of the outgoing 
wave, the complete frequency dependence of (\ref{eq:emrategen}) 
can be accounted for 
qualitatively~\cite{strominger97a,mathur97b,gubser,cl97b}.  
Thus, for large partial wave numbers, the microscopic description in 
terms of an effective string model accounts for the emission 
rates {\it in  an arbitrary black hole background}.

An important unresolved problem remains the calculation of the 
overall numerical coefficient in  (\ref{eq:emrategen}). However, 
this issue is not specific to the non-extreme case.
It is expected that this coefficient is calculable 
in the BPS-limit and that it agrees with the classical result.
If this is borne out it is will also be possible to model the 
general non-extreme black hole.

The vertex operators (\ref{eq:vertex}) are fermionic when $n$ is 
odd. Therefore  the phase-space factors associated with the 
initial state are expected to be of the Fermi-Dirac type. In this case the 
Green's functions  (\ref{eq:green}) are proportional to gamma functions 
with arguments whose real parts  are half-integral, which indeed give 
factors of the Fermi-Dirac form 
$(e^{\omega/2T}+1)^{-1}$~\cite{strominger97a}. This is an 
interesting example where the black hole ``looks'' like a 
string with fermionic degrees of freedom.

\section{Conclusion}
\label{sec:conclusion}
We conclude by summarizing the accomplishments and the shortcomings
of the effective string model for non-extreme black holes,
starting with the former:
\begin{itemize}
\item
The model relates the thermodynamic variables in the effective (weakly coupled)
 string description
directly to geometrical features of the black hole space-time.
\item
The extreme and near-extreme limits  are in agreement with  the model. This result
provides a 
strong motivation  that a general non-extreme black hole  can be modeled by an
effective string model as well.
\item
Two-body processes can be accounted for quantitatively.
Specifically this  result gives a microscopic interpretation of the universal 
low-energy absorption cross-section.
\item
Many-body processes can be understood qualitatively.
\end{itemize}

Despite these successes it is appropriate to conclude 
with some caution. The 
understanding of non-extreme black holes presented here leaves much 
room for improvement:
\begin{itemize}
\item
The detailed  connection with a specific
fundamental string theory, along with the detailed description of the
underlying SCFT, is not clear.
\item
The description of the string spectrum is limited to the black hole
regime because we employ an approximate notion of an effective level.
\item
Current models of black holes in string theory are not sensitive to 
the nontrivial causal structure of black hole space-times. 
\end{itemize}
It is not clear whether these obstacles can be overcome with
further developments of the ideas and techniques that are presently known.

\vspace{1.cm}

{\bf Acknowledgment}
We would like to thank G. Gibbons, S. Gubser, G. Horowitz, S. Mathur,  
E. Verlinde, and H. Verlinde for discussions. The work was supported by 
DOE grant DE-FG02-95ER40893 and NATO collaborative grant
CGR 949870 (M.C.). We would like to thank the Aspen Center for
Physics  (M.C.) and  NORDITA  (F.L.) for 
hospitality while the manuscript was being  prepared.

\end{document}